# Universal Boolean Logic in Cascading Networks

Galen Wilkerson and Sotiris Moschoyiannis

University of Surrey, Guildford, Surrey, GU2 7JP, UK
g.wilkerson@surrey.ac.uk, s.moschoyiannis@surrey.ac.uk

**Abstract.** Computational properties of networks that can undergo cascades are examined. It is shown that universal Boolean logic circuits can be computed by a global cascade having antagnistic interactions. Determinism and cascade frequency of this antagonistic model are explored, as well as its ability to perform classification. Universality of cascade logic may have far-reaching consequences, in that it can allow unification of the theory of computation with the theory of percolation.

## 1 Introduction

Cascades are ubiquitous phenomena, found in social decision making, information diffusion, disease spreading, neuronal firing, and many other biological, social, chemical, and physical systems [5]. A classic example is a sand-pile, gradually built up grain by grain, until its sides reach a critical slope, undergoing a phase transition, where it tends to experience large avalanches (cascades) [3]. Another example is a neuron. When firing at a certain rate, the neuron may act alone, but when it fires at a critical rate, it may trigger a large cascade, causing many other neurons to fire [4], possibly optimizing information processing [12]. These models are also related to *percolation theory* [1].

The ubiquity of cascades in many naturally-occurring systems is compelling, both for scientific understanding and as an important mechanism to advance computing. As Moore's law is challenged by physical limitations, alternative avenues to speed up computing are being investigated. Organic computing has been studied due to its great efficiency and adaptability [8]. In fact, when we look at the ability of the brain to learn quickly and perform many highly complex functions in parallel, we see that there is a huge efficiency gap between this and the fastest modern computer [10].

Thus, learning in naturally-occurring networks can be compared to 'deep learning' by artificial neural networks [10], or studied as a problem of control of complex systems [9].

The McCulloch-Pitts Linear Threshold Unit and early neural networks were motivated by an interest in mimicking brain function. However, it seems that in our modern era of deep-learning, the focus has been to use stochastic gradient descent and back-propagation because they are good engineering that works, but perhaps digress from the initial inspiration of the brain [10].

---

1 *Percolation* is a classic model in physics and graph theory describing the sudden appearance of a giant component as a function of connection probability in Erdos-Renyi graphs. It can also be related to flow of liquid through a porous medium.

Meanwhile, machine learning is pre-dated by and based upon formal theories of computation, having a long history, starting with logic and computing with Boole and Babbage and famously advanced with computability and circuit complexity by Turing, Shannon, and others, describing how simple systems can be combined to powerful effect [11].

Thus, the simplicity of the GCM, a networked form of the Linear Threshold Model [6], and its Boolean construction (edges and node states), along with its possibility to encode logic circuits (shown here), and its critical cascade behavior, make it a compelling framework to explore the relationships between theories of computation and theories of criticality and cascades. Both in computer science and physics it is clear that powerful and significant large-scale behaviour can emerge from the simplest of models.

The remainder of this paper is organized as follows: In section 2 we examine some characteristics of the simple Global Cascade Model (GCM) [13], and show that it can compute Boolean circuits. In section 3, we introduce a small modification to create an Antagonistic Global Cascade Model (AGCM) having negative interactions, and in section 4 show that it can compute functionally-complete (universal) Boolean logic circuits. In section 5, we also show that the GCM can reach deterministic final states, but the AGCM's final state is not always deterministic. In section 6 the AGCM and GCM are also shown to have a complementary cascade frequency distribution. Finally, in section 7 there is a discussion of a few issues and future directions.

## 2 The Global Cascade Model (GCM)

First, let's briefly review the GCM (on a finite graph) [13]:

The GCM runs as follows:

- Create an Erdos-Renyi random network, $G(N,p)$, with a fixed number of nodes $N$ and edge probability $p$.

- Assign a random threshold value $\phi \sim U[0,1)$ to each node.

- All nodes are marked as **unlabeled**.

- Randomly choose a small fraction ($\Phi_0 \ll 1$) of the graph's nodes to be seeds, mark them **labeled**.

- Run the simulation by asynchronously checking each unlabeled node, applying the threshold function

$$f(u, \nu, \phi) = \begin{cases} \textbf{label u} & \nu \geq \phi \\ \textbf{no change} & \nu < \phi \end{cases} \qquad (1)$$

for node $u$ and $\nu = \dfrac{N_L(u)}{deg(u)}$, the fraction: number of node $u$'s neighbors that are labeled $N_L(u)$, over its degree $deg(u)$.

- Stop after all unlabeled nodes have been examined without any change in labeling. The *cascade size* is the fraction of nodes ($\Phi$) that are labeled. A *global cascade* is said to occur if the cascade size exceeds a predetermined fixed fraction of the network (e.g. $\Phi \geq \frac{1}{2}$).

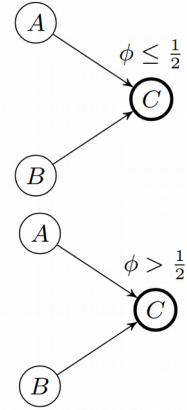

```
A|B| ν  |C
0|0|0/2|0
0|1|1/2|1   ≡ OR
1|0|1/2|1
1|1|2/2|1

A|B| ν  |C
0|0|0/2|0
0|1|1/2|0   ≡ AND
1|0|1/2|0
1|1|2/2|1
```

*Figure 1: A graph focusing on node C's behavior from neighbors (inputs) A and B under the original labeling rule (Eq. 1). Top: Truth table for OR and corresponding cascade network when node C's threshold $\phi \leq \frac{1}{2}$. Bottom: Truth table for AND and cascade network when C's $\phi > \frac{1}{2}$. Networks drawn as directed to show flow of information.*

Nodes in the GCM behave like logic gates. Taking a look at one unlabeled node (node *C*, Fig. 1) in a graph with two neighbors, we observe that when *C*'s threshold $\phi \leq \frac{1}{2}$, considering nodes *A* and *B* as inputs, and using the labeling rule (Eq. 1), node *C* behaves like a logical **OR**. Similarly, when $\phi > \frac{1}{2}$, node *C* behaves like logical **AND**.

Generally, for *k* inputs, a node behaves like multi-input OR for any $\phi \leq 1/k$, and multi-input AND for any $\phi > (k-1)/k$. For $1/k < \phi \leq (k-1)/k$, a node behaves like a threshold logic unit [10].

As we see in the truth tables (Fig. 1), this model can only carry out *monotonically increasing* logical functions such as the identity, AND, or OR [10]. That is, given two n-dimensional points $x = (x_1, ..., x_n)$ and $y = (y_1, ..., y_n)$, a function *f* is monotonically increasing if $f(x) \geq f(y)$ when the number of 1s in *x* is at least the number of 1s in *y*.

## 3 The Antagonistic Global Cascade Model (AGCM)

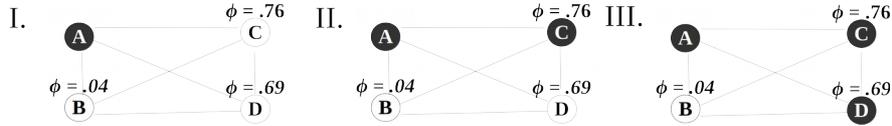

*Figure 2: An example of a cascade using the antagonistic model's complement rule (Eq.2), starting at seed node A (step I) and proceeding to nodes C (step II) and D (step III).*

We now introduce the Antagonistic Global Cascade Model (AGCM) having *antagonistic* interactions, by making a single modification to the GCM.

We construct a new *complement labeling rule* by simply reversing the inequalities of Eq. 1:

$$\neg f(u, \nu, \phi) = \begin{cases} \textbf{label u} & \nu < \phi \\ \textbf{no change} & \nu \geq \phi \end{cases} \quad (2)$$

again for node *u* and *v* the fraction of neighbors labeled, as above.

Other than this rule, the operation of the AGCM is the same as the GCM. The new model's cascade action can be observed in Figure 2.

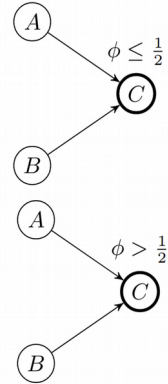

| A | B | $\nu$ | C |
|---|---|-----|---|
| 0 | 0 | 0/2 | 1 |
| 0 | 1 | 1/2 | 0 | ≡ **NOR**
| 1 | 0 | 1/2 | 0 |
| 1 | 1 | 2/2 | 0 |

| A | B | $\nu$ | C |
|---|---|-----|---|
| 0 | 0 | 0/2 | 1 |
| 0 | 1 | 1/2 | 1 | ≡ **NAND**
| 1 | 0 | 1/2 | 1 |
| 1 | 1 | 2/2 | 0 |

*Figure. 3. Node C's behavior under the complement labeling rule (Eq. 2). Top: Truth table for NOR and corresponding cascade network whenever node C's $\phi \leq$ ½. Bottom: Truth table for NAND and cascade network whenever C's $\phi >$ ½.*

The new labeling rule takes the logical complement of the original Boolean circuits, computing NOR in the place of OR, and NAND in the place of AND (Fig. 3), for identical $\phi$ values. The multi-way logic mentioned above is also complemented in the same way.

NAND and NOR, are *monotonically decreasing* Boolean functions (Increasing *true* values input decreases *true* values output), as can be seen in the truth tables in Fig. 3.

## 4 Boolean Circuits and Functional Completeness

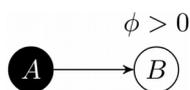

*Figure. 4.: Trivially, we can construct NOT for a node with a single neighbor and any $\phi >0$ value. This can be appended to a NAND or NOR circuit to give AND or OR.*

After a cascade is run (in either model), when all unlabeled nodes have been examined without any change, the resulting logic circuits have no memory or time component, so are called *combinational logic circuits* that compute *straight-line programs* [11].

On a practical level, it is fairly straightforward to wire together Boolean networks. Logical NOT can be trivially constructed for any $\phi > 0$, (Fig. 4) to build up the equivalence of Boolean circuits and Boolean algebra, including NAND and NOR, and their complement, AND and OR, and any algebraic combination [10].

Recall that *functionally complete* logic circuits can be created by composition of *solely* **NAND** or *solely* **NOR** operations, either of which form a *universal basis* [11]. This means that any Boolean circuit can be created using a network using this complement labeling rule.

For example, in Fig. 5, we have shown how one particular random AGCM is able to carry out the logic of a half-adder [2] if the unlabeled nodes are examined in one of several orders, but not necessarily all possible orders (discussion below). *A network composed of nodes using the GCM labeling rule* (Eq. 1) *could not have performed this calculation in a cascade.*

We note now that the computational power of such networks is in their *scaling* of universal Boolean logic. That is, functional completeness in a large network allows us to build any logical operation needed, including adding, subtracting, multiplying, dividing, and much more complex operations, including certain kinds of computer programs and deciding classes of formal languages [11].

---

2 A *half-adder* is a Boolean circuit that adds two bits and outputs a sum and carry bit, much like standard base 10 addition.

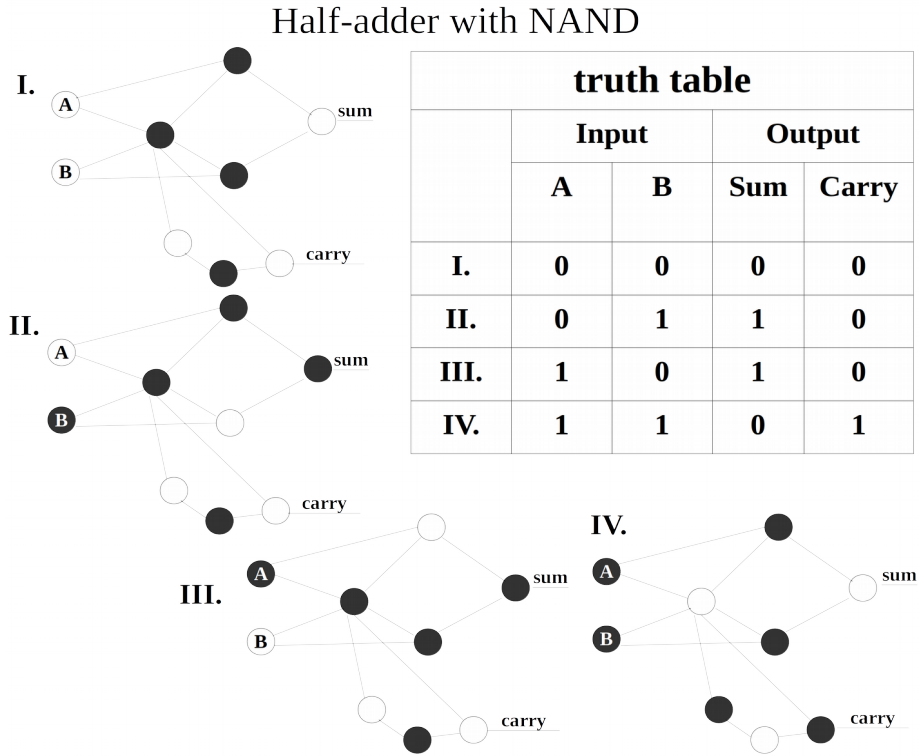

*Fig. 5. A half-adder implemented by a cascade network having the complement labelingrule. Here all nodes have been assigned $\phi$ values equivalent to NAND. This is a very small example of the power of the functionally-complete Boolean logic that can becomputed by these networks.*

## 5  Determinism

As we have seen, the GCM's labeling rule is monotonically increasing. Therefore, the original labeling rule leads to convergence to the final state of the network as unlabeled nodes are examined, and this final state is *deterministic* (unique). The proof is as follows.

First we show that no labeling can exclude another labeling in the original GCM:

**Theorem 1.** *Labeling a particular node in the GCM cannot exclude labeling another node.*

*Proof.* Suppose, at a particular time-step, that examining some node *u* in the network leads to labeling *u*, and this labeling excludes the future labeling of some node *v* in the network. This contradicts the labeling rule (Eq. 1), which is monotonically increasing in its inputs. □

After a node has become labeled, it is sufficient to examine all unlabeled nodes $U \leq N$ only once to determine whether the cascade has completed:

**Lemma 1.** *After a change in labeling of the GCM, it is sufficient to examine each unlabeled node once to determine whether the cascade has completed.*

*Proof.* Suppose that at a particular time step $t$, node $u_1$ becomes labeled. Now suppose that after $u_1$ has become labeled we have to examine some unlabeled node $u_2$ more than once to determine whether $u_2$ has become labeled. Letting $t_1$ be the first time we examine $u_2$ and $t_2$ be the second time we examine $u_2$, this means we have labeling rule $f_{t_1}(u_2, \eta, \phi) \neq f_{t_2}(u_2, \eta, \phi)$, for the same inputs. This contradicts the labeling rule $f$ (Eq. 1). □

Also, a cascade cannot remain blocked if there is an unlabeled vulnerable node, simply because this node has not been examined:

**Theorem 2.** *A cascade cannot be blocked indefinitely due to not examining a particular unlabeled node.*

*Proof.* Suppose node $v$ is the only node that will allow a cascade to continue. At a particular time step $t$, the probability of not examining node $v$ is $p_{v'} = \frac{U-1}{U} < 1$, for $U$ the number of unlabeled nodes to be examined (by Lemma 1). As each unlabeled node is examined at random,

$$\lim_{t \to \infty} p_{v'} = \left(\frac{U-1}{U}\right)^{\infty} = 0$$

Therefore eventually node $v$ will be examined and the cascade will continue. □

As a result, all cascades in the original GCM must eventually converge to the same final state:

**Theorem 3:** *The final state of the GCM is unique.*

*Proof.* Suppose that for $G(N, p, s, \phi, f)$, the random graph $G$ having $N$ nodes, edge probability $p$, seed node(s), phi values $\phi$, and labeling rule $f$, the GCM cascade operation is run separately from start to finish two times, and converges to two different final states.

This means that for one of the two runs, either A. labeling one node excluded labeling another node, or B. the cascade was blocked because a particular unlabeled node was never examined. Case A contradicts Theorem 1, and B contradicts Theorem 2. □

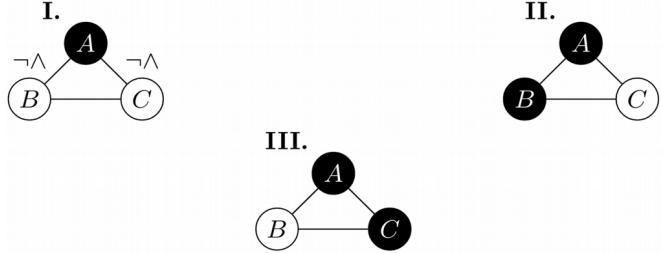

*Fig. 6: The complement labeling rule creates order-dependence. In a small fixed network, nodes B and C have φ equivalent to NAND. I. Start with seed node A labeled. The final state of the network depends on whether we examine II. B first or III. C first.*

This leads us to an interesting characteristic that arises when we use the AGCM's complementary labeling rule – the final state of the cascade becomes dependent on the order in which we examine the unlabeled nodes, and therefore unlike the GCM, is *non-deterministic* (non-unique). Observe the triangular graph in Fig. 6, where nodes *B* and *C* have $\phi$ threshold values equivalent to NAND. Starting with seed-node *A*, if we first examine node *B* (top-right), we reach a final state with node *B* labeled and node *C* unlabeled. However, *B*'s and *C*'s rules are identical, so we have broken symmetry. A similar behavior can be observed in graphs without a cycle.

## 6    Cascade Frequency

It appears that the modality of the AGCM's cascade frequency as a function of the average degree *z* is complementary to the cascade frequency of the original GCM (See [13], Fig. 2b). An intuition is that in the AGCM network cascade, an increase in the number of labels tends to have an antagonistic effect on the number of subsequent labels. Here we have used a uniform $\phi^* = 0.18$ for all nodes, as in the original work [13], for graphs having 100,000 nodes over 100 realizations, for the integer-valued mean degree $z \in [1,10]$, to calculate the frequency that the cascade size exceeds median cascade size (Fig. 7). While in the GCM, the largest cascade frequency occurred in ($2 \leq z \leq 6$), here we see the smallest cascade frequency in that region.

Thus, it seems that following AGCM's complement rule means that cascades occur almost exactly when they do not in the original GCM. While in the original model, a very sparse or disconnected network can easily block or cut off a cascade, in the complement label case a sparsely connected network will tend to create more labeling, as we see by the left-hand mode, for $z \leq 2$ (Fig. 7).

In the region around $z = 5$, we see low cascade frequency (Fig. 7). In this region, the graph may still be tree-like [13], with low clustering coefficient $\left(\overline{C} \propto \frac{z}{N-1} \approx 0.00005\right)$, so that seeds largely reach nodes from only one neighbor, but since we have chosen $\phi^* = 0.18$, nodes are not vulnerable, since $v = 1/5 > 0.18 = \phi^*$.

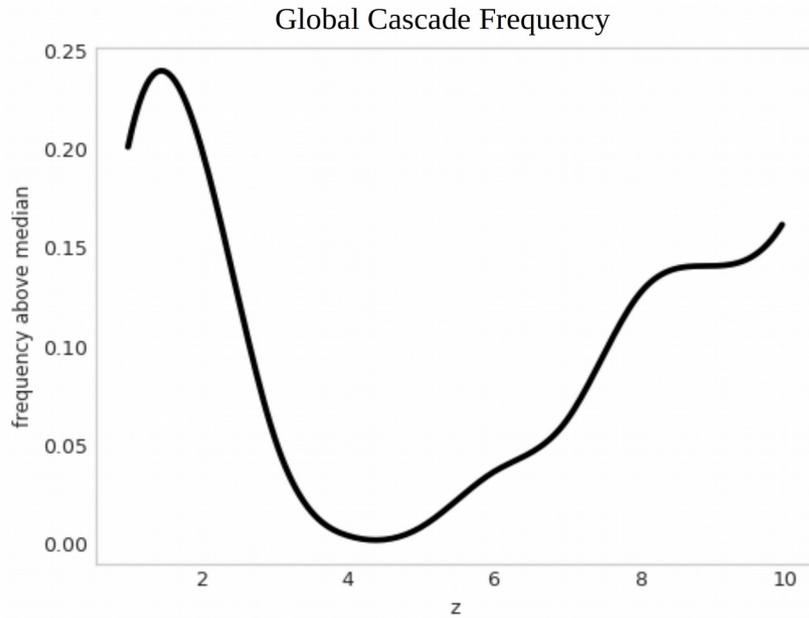

*Fig. 7. Bi-modality in the global cascade frequency using the complement labeling rule, here on a network with 10000 nodes over 100 realizations per average degree z, and a uniform $\phi^* = 0.18$.*

In the original GCM model, as $z$ increases, densely connected unlabeled nodes tend to dilute the effects of incoming labeled nodes, which are often unable to overcome thresholds $\phi$, driving down the cascade frequency. In the right-hand side ($z \geq 8$) of the AGCM complement case (Fig. 7), unlabeled dense networks tend to be easily influenced by incoming labels, increasing labeling until negative feedback (antagonism) discourages further labeling, as seems indicated by the lower peak in the right-hand mode.

## 7   Discussion

Here we address a few points that may have occurred to the reader.

One might rightly ask how networks with solely negative threshold rules could come about, where examples of them could be found, and whether it is reasonable to study them. Antagonistic interactions have been investigated in the area of social balance [7], consensus [1], and multi-edge graphs [14].

Above, we have discussed cascading networks as Boolean circuits. However, we can also think of them as performing a *classification* as in *machine- or deep-learning*. The AGCM labeling rule (as well as the original GCM rule) for node $u$ can be written in the form of a *threshold logic unit* used in neural networks [10]:

$$g(u, \bar{x}, \phi) = \begin{cases} \textbf{label u} & (\overline{w} \cdot \overline{x})/\|\overline{x}\| < \phi \\ \textbf{no change} & otherwise \end{cases}$$

where $\|x\|$ is the L1-norm (number of ones) of the Boolean input vector *x* and *w* = $[1,1,1,\ldots,1]^T$.

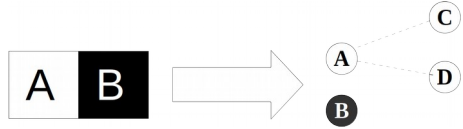

*Fig. 8: A thought experiment showing why the original monotonically increasing GCM rule cannot distinguish between only pixel B being black, or both pixels being black. The construction of a network (nodes C and D) attached to A cannot detect when A is unlabeled (white).*

Further on classification, at first impression it may seem that the original GCM may compute a NOT simply by the absence of a cascade (and thereby classify 'cat' or 'not cat', for example). However, this is not the case. For example, as in (Fig. 8) we may construct a simple GCM taking an image as its possible seed nodes, and attempt to detect when only the right-hand pixel (B) is black. (Here, black = 'labeled', white = 'unlabeled' as elsewhere.) However, since the GCM rule is monotonically increasing, it is impossible to detect when the left-hand pixel (A) input is unlabeled, thereby possibly obtaining many false positive results. However, the AGCM, having NOT, can resolve this situation.

Similarly, the classic XOR problem [10] cannot be resolved by the GCM (since the composition of monotonically increasing functions is monotonically increasing) but can be resolved by the AGCM, as we see in the sum column of the half-adder truth table (Fig. 5).

Finally, we note now that the computational power of such networks is in their *scaling*. That is, functional completeness in a large network allows us to build *any logical operation*, including adding, subtracting, multiplying, dividing, and very complex operations, including certain kinds of computer programs and classes of formal languages [2]. If we consider a very large network (the brain has ~ $10^{11}$ neurons and ~ $10^{14}$ synapses), it is hard to imagine the capabilities.

A large number of topics remain to be investigated, unfortunately relegated to 'future work'. These are so many it is only possible to list them briefly:

Some of these topics are theoretical: Investigating the existence of a critical threshold in the AGCM model, and finding a closed-form expression for it; How to train or control the network toward criticality for certain inputs; Analysing both the GCM and AGCM as dynamical systems, understanding their correlation length and time, stability and fixed points, as well as convergence and accuracy evolution of the logic circuit; understanding how input size, cascade size and criticality relate to circuit size and depth and information-theoretic measures; understanding how to extract outputs from the network, using either cascade size or individual nodes; studying how a mix of nodes having one or the other labeling rule can be functionally complete; the relationship between tree-like topology and logic that can be computed; how criticality in these simple models relates to optimal functionality in neuronal networks of the brain [12]; the formal language (called *AC* for Boolean circuits) that can be computed by these cascades, and the probabilistic relationship to criticality; and whether these Boolean cascades can be considered in quantum computing using qubits.

Other topics are more applied: How to build such networks in hardware; consideration of real-world network topologies (rarely Erdos-Renyi); observing and measuring such cascades 'in the wild' in social, neuronal, or other systems.

## 8 Conclusion

We have shown that a simple cascading network model can compute Boolean circuits, and that a cascading network having antagonistic interactions can compute universal, functionally-complete Boolean logic. We have also shown that, although cascades over positive interactions can reach deterministic final states, antagonistic interactions may not. The antagonistic model has a complementary cascade frequency distribution to the original GCM, and may perform classification, as in machine learning.

This research lets us begin to understand relationships between theories of ubiquitous cascades in nature and Boolean circuits that are fundamental to computing.